\documentclass[floatfix,aps,prd,showpacs,amsmath,nofootinbib,preprintnumbers]
{revtex4}
\usepackage{graphicx}
\usepackage{colordvi}

\def\be{\begin{equation}}
\def\ee{\end{equation}}
\def\ba{\begin{eqnarray}}
\def\ea{\end{eqnarray}}
\begin{document}
\title{Evolution of Large Scale Curvature Fluctuations \\ During the 
Perturbative Decay of the Inflaton}
\author {Fabrizio Di Marco}\email{dimarco@bo.infn.it}
\affiliation{Dipartimento di Fisica, Universit\`a degli Studi di Bologna and
  INFN Via Irnerio, 46-40126, Bologna, Italy}
\author{Fabio Finelli}\email{finelli@iasfbo.inaf.it}
\affiliation{INAF/IASF-BO,
Istituto di Astrofisica Spaziale e Fisica
Cosmica di Bologna \\
Via Gobetti 101, I-40129 Bologna - Italy}
\affiliation{INAF/OAB, Osservatorio Astronomico di Bologna,
Via Ranzani 1, I-40127 Bologna -
Italy}
\affiliation{INFN, Sezione di Bologna,
Via Irnerio 46, I-40126 Bologna, Italy}
\author{Alessandro Gruppuso}\email{gruppuso@iasfbo.inaf.it}
\affiliation{INAF/IASF-BO,
Istituto di Astrofisica Spaziale e Fisica
Cosmica di Bologna \\
Via Gobetti 101, I-40129 Bologna - Italy}
\affiliation{INFN, Sezione di Bologna,
Via Irnerio 46, I-40126 Bologna, Italy}
\date{\today}
\begin{abstract}
We study the evolution of cosmological fluctuations during and after 
inflation driven by a scalar field coupled to a perfect fluid through a 
friction term. During the slow-roll regime for the scalar field, the perfect 
fluid is also frozen and isocurvature perturbations are generated.
After the end of inflation, during the decay of the inflaton, 
we find that a change in the observationally relevant large scale curvature 
fluctuations is possible. 
\end{abstract}

\maketitle



\section{INTRODUCTION}\label{Introduction}



The inflationary paradigm provides an explanation for the 
large-scale curvature fluctuations seen in the pattern of anisotropies of 
the cosmic microwave background and of the large scale structure 
\cite{books}.
In an inflationary scenario driven by a single
scalar field, large-scale curvature perturbations are generated by
the amplification of quantum fluctuations of the inflaton field
during the accelerated expansion era and remain constant until their 
reentry in the Hubble radius in the absence of post-inflationary changes. 

In the context of cold inflation, the release of entropy is 
left for the post-inflationary evolution, 
during which the inflaton decays into other 
intermediate fields or directly into the matter our present universe is made of.

The possibility of a change in the amplitude of large scale curvature perturbations
during preheating has been widely investigated 
\cite{bassett1,bassett2,FB_PRL,FB_PRD,JS,ivanov,FK}.
Such a change is possible due to the coupling of isocurvature and curvature perturbations
on large scales \cite{bardeen,mollerach}.
For additional scalar fields coupled to the inflaton, the possibility of 
a post-inflationary amplification of curvature perturbations is very model dependent, but possible and 
strongly related to the spectrum of the decay products generated during inflation.

Although a post-inflationary change of large scale curvature perturbations during 
reheating might be an unexpected twist for successfull single 
field inflationary models, it may be useful for the single field models 
whose parameters clash with observations.
Along this line, after it was shown that curvature perturbations can indeed 
change during preheating, it was proposed that large scale curvature 
perturbations may be seeded by light scalar fields during 
reheating \cite{DGZ}.
The proposal to seed curvature perturbations at reheating 
requires three components, according to \cite{DGZ}: 
the inflaton, the decay products 
(taken as a perfect fluid) coupled to the inflaton 
and a light modulus on which the decay rate 
depends. 

Although a lot of attention has been paid to system of 
scalar fields, also interacting through the kinetic terms 
\cite{DFB,koshelev}, here instead we analyze the
evolution of cosmological perturbations in an inflationary scenario driven 
by a standard scalar field coupled to a perfect fluid (with equation of state $p_F =\omega_F \rho_F$) through a friction term $\Gamma$. 
This setting is motivated from the old theory of reheating 
\cite{ASTW,DL,AFW} to describe the decay of the inflaton into matter 
after the accelerated stage, but we believe
it is also interesting for several reasons. First, 
fluctuations of a scalar field with a non-vanishing potential 
are different from those of a perfect fluid. 
As a second point, it is conceivable to explore the consequences of a coupling for the inflaton 
not dictated by its interaction with another scalar field.
Third, the interaction between the inflaton and the perfect fluid 
is invisible to gravity, in contrast with what happens to scalar fields 
where a term $g^2 \phi^2 \chi^2$ appears as part of the potential driving the 
inflationary stage. 
For these three reasons duplication of the results of two-field inflationary
models is not expected.
This formalism is also suitable for investigating the 
prediction of warm inflation \cite{WarmInflation,DeOliveira,leefang}, 
in which the dissipative term $\Gamma$ is much 
larger than the Hubble parameter during inflation. 
Note that here the numerical value for the ratio 
$\Gamma/H$ is left free in this paper.

We stress that the main goal of this paper is to study cosmological 
perturbations for an inflaton coupled to a perfect fluid 
{\em during and after} inflation. To our knowledge, 
in previous studies of cosmological 
perturbations in warm inflation the post-inflationary stage was not 
considered \cite{DeOliveira,leefang}.
Apart from warm inflation, 
this setting has been applied only after 
inflation to describe the reheating stage with a decay rate computed 
by quantum field theory methods. 
However, the inflationary geometric amplification of 
zero-point fluctuations applies to any component present 
in the Lagrangian. If we think about the inflaton 
coupling to a perfect fluid as an effective way to describe its couplings 
to other degrees of freedom present in the Lagrangian, 
it is conceivable to consider the perfect fluid contribution also during inflation.   

The outline of paper is as follow. In section II and III
we present the governing equations for the background and scalar 
perturbations in the longitudinal gauge, respectively.
In section IV we present the evolution of scalar perturbations 
in the uniform curvature gauge (UCG henceforth), from which is easy to obtain the coupled 
equations for the Mukhanov variable 
(the gauge invariant scalar field fluctuation \cite{mukhanov}) 
and the Lukash variable (the gauge invariant fluid 
fluctuation \cite{lukash}).
The numerical analysis presented is based on this last set of equations.
In section V we conclude.

\section{Background Evolution}
In this section we present the 
background equations for the scalar field plus perfect fluid and several 
background quantities, as the speed of sound for the various components.
The equations of motion for the scalar field and the fluid we consider 
are the following:
\be
\ddot{\phi} = - 3 H \dot\phi - \Gamma \dot\phi - V_{\phi} \label{motophi}
\ee
\be
\dot\rho_F = - 3 H (1 + \omega_F)\rho_F + \Gamma \dot\phi^2\, ,
\label{motoF}
\ee
where $H \equiv \dot a / a$, $w_F$ is the state parameter of the fluid 
- which we consider non negative - $\Gamma$ is the friction coefficient - 
which can depend on any other quantity. 
Note that this coupling has a fully covariant description \cite{leefang} 
and that other coupling of the inflaton 
to a perfect fluid may be considered \cite{DGZ}.
The above equations correspond to: 
\be
\dot \rho_i + 3 H (1 + w_i) \rho_i = X_i \,,
\ee
where $i=\phi, F$ and $X_\phi = - X_F = -\Gamma \dot \phi^2$.
The Einstein equations are:
\be
H^2 = \frac{8\pi G}{3} \left[ \frac{\dot\phi^2}{2} + V(\phi) + \rho_F \right]
\equiv \frac{8\pi G}{3} \rho_{\rm tot} \,,
\label{hubble}
\ee
\be \dot H = - 4 \pi G \left[ \dot\phi^2 + \rho_F(1+\omega_F) \right] 
\equiv - 4 \pi G \left( \rho_{\rm tot} + p_{\rm tot} \right) \equiv 
- 4 \pi G \rho_{\rm tot} \left( 1 + w_{\rm tot} \right)
\,,
\label{hubbledot}
\ee
where it is clear that the dissipative term $\Gamma$ does not enter 
explicitly in the Einstein equations. The dissipative term $\Gamma$ 
slows down the inflaton in a slow-roll regime and damps it into the 
fluid during the oscillatory stage 
when the value of the Hubble rate $H$ drops below $\Gamma$.

For future convenience we define the parameter $\omega_{\phi}$ as:
\be 
\omega_{\phi} 
= \frac{p_ {\phi}}{\rho_{\phi} } =  \frac { \dot\phi^2
  /2 - V (\phi) }{ \dot\phi^2 /2 + V (\phi)} \label{omegaphi} \ee
and the sound of speed associated to scalar field as:
\be  c^2_{\phi} \equiv \frac{\dot p_ {\phi}}{\dot\rho_{\phi} } = 1 +  \frac{2
  V_ {\phi}}{3H\dot\phi (1 + r) } \,. \label{soundphi} \ee
where $r = \Gamma / (3 H)$. This last formula is the 
generalisation of Eq. (6) of \cite{Bartolo-Corasaniti}
to the case $ r \neq 0$. We also define 
the total speed of sound:
\be  c^2_{\rm tot} \equiv
\frac{\dot p_{\phi} + \dot p_F}{\dot\rho_{\phi} + {\dot\rho_F}} =  
\omega_{\rm tot} - \frac{\dot w_{\rm tot}}{3 H ( 1 + w_{\rm tot})}
= c^2_{\phi} +
\frac{p_F +\rho_F - r (p_{\phi}  +\rho_{\phi} )}{p_{\rm tot} +\rho_{\rm
    tot}}\left(\omega_F - c^2_{\phi} \right)\, .
  \label{soundtot} \ee

When the inflaton slow-rolls, the following approximations hold:
\be
\dot\phi \simeq - \frac{V_{\phi}}{3H(1 + r)}
\label{slowphi} \ee
\be  \rho_F \simeq \frac{\Gamma \dot\phi^2}{3H(1+\omega_F)}
= \frac{r \dot\phi^2}{(1+\omega_F)}  
\label{slowF}
\, .\ee
It is clear that the perfect fluid also remains almost 
frozen because of the coupling to the inflaton, i.e. $\dot \rho_F \simeq 0$. 
The (cosmic) time derivative of the Hubble distance is:
\be 
\epsilon \equiv - \frac{\dot H}{H^2} = 
\epsilon_\phi + \epsilon_F = \frac{4 \pi G \dot \phi^2}{H^2} + 
\frac{4 \pi G \rho_F (1 + w_F)}{H^2}
\simeq \epsilon_\phi (1+r)
\label{epsilon}
\ee
where $\simeq$ denotes the 
validity during slow-roll.

We note that this coupling of the inflaton $\phi$ to a perfect fluid is 
inequivalent to the coupling to another massless scalar field $\chi$ 
in a quartic way $g^2 \phi^2 \chi^2$. If we study such a system:
\be
H^2 = \frac{8\pi G}{3} \left[ \frac{\dot\phi^2}{2} + V(\phi) + 
\frac{\dot\chi^2}{2} + g^2 \frac{\phi^2 \chi^2}{2} \right] \equiv 
\frac{8\pi G}{3} \left[ \rho_\phi + \rho_\chi \right]
\,.
\label{hubble_scalar}
\ee
By considering $\rho_\chi = \dot\chi^2/2 + g^2 \phi^2 \chi^2/2$, 
the structure of the corresponding interacting term $X_\phi$ is different 
from the one in Eq. (\ref{motoF}). 

\begin{figure}
\begin{tabular}{cc}
\includegraphics[scale=0.54]{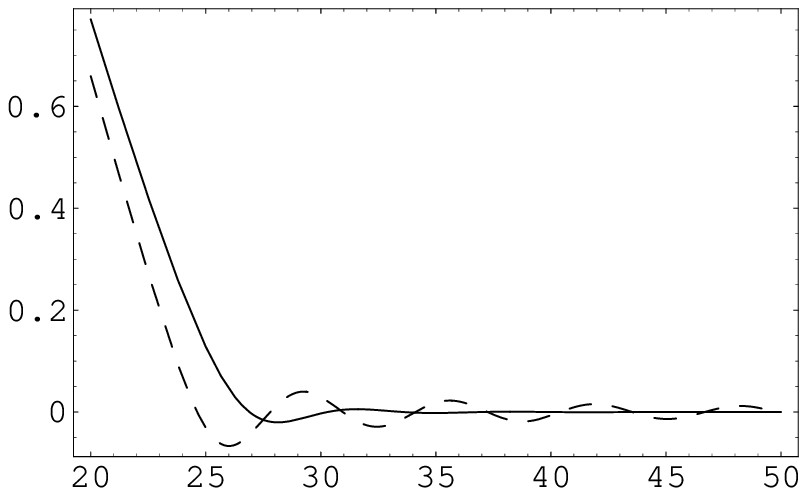}
\includegraphics[scale=0.54]{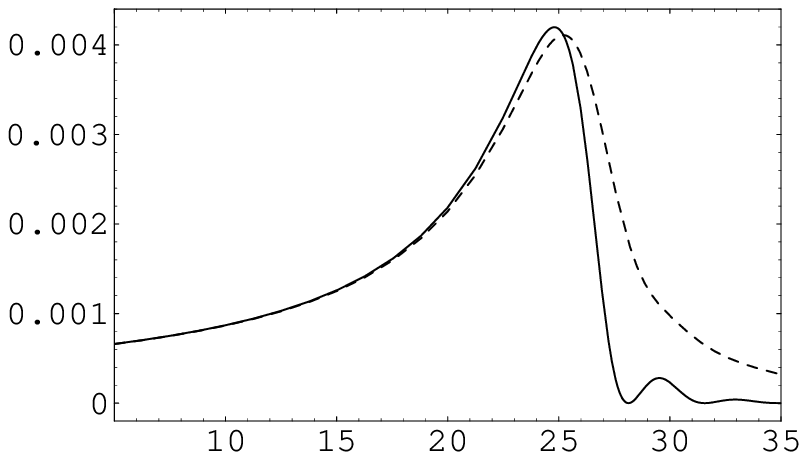}
\includegraphics[scale=0.54]{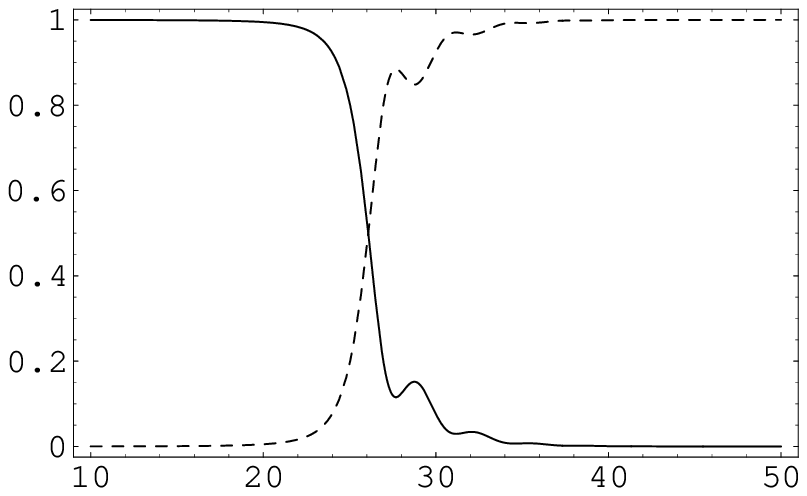}
\includegraphics[scale=0.54]{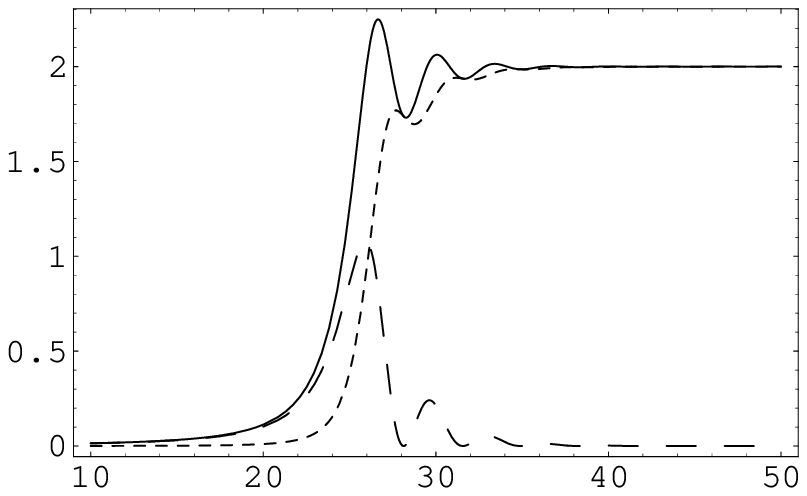}
\end{tabular}
\caption{Typical evolution of background quantities from inflation to 
radiation ($w_F=1/3 \,, \Gamma/m =0.5$) for $V (\phi) = m^2 \phi^2/2$ 
as function of an adimensional 
cosmic time $m \, t$. On the left panel we test validity of Eq. (\ref{slowF}):
we show the 
evolution of $\phi \sqrt{G}$ in the coupled 
(solid line) and uncoupled (dashed line) case. Note how Eq. (\ref{slowF}) 
is a good approximation during inflation.
In the second panel the time evolution of 
$\Gamma G {\dot \phi}^2 / (3 H m^2)$ (solid line) 
and $4 \pi G \rho_F (1+w_F)/m^2$ (dashed line). 
The time evolution of $\rho_\phi / \rho_{\rm tot}$ (solid) and 
$\rho_F / \rho_{\rm tot}$ (dashed) is displayed in the third panel.
On the right panel it is shown the 
evolution of $-\dot H/H^2$ (solid line) and its contribution from 
the scalar field $4 \pi G {\dot \phi}^2/H^2$ (long-dashed line) and from 
radiation $4 \pi G \rho_F (1+w_F)/H^2$ (short dashed line).}
\label{fig1}
\end{figure}

As a final comment of this section, it is useful 
to compare this setting with the $\Gamma=0$ case, which 
is very similar to the single inflaton case since 
$\rho_F \propto a^{-3 ( 1 + w_F)}$ is rapidly washed out for $w_F >0$.
The dissipative term $\Gamma$
slows down the inflaton during the slow-roll regime and damps its amplitude 
during the oscillatory stage. As an example, we show the left panel of 
Fig. \ref{fig1} the evolution of a massive inflaton - 
$V(\phi) = m^2 \phi^2 / 2$ - with $\Gamma = 0.5 m$. See the caption 
for the evolution for other relevant background quantities.

\vspace{0.7cm}

\section{Scalar Perturbations in the Longitudinal Gauge}

The scalar perturbations around a flat Robertson-Walker metric are:   
\be
ds^2=-(1+2 \alpha)dt^2-a \beta_{,i} dt dx^i+a^2 \left [
\delta_{ij}(1 - 2 \phi)+2 \gamma_{,ij} \right ]dx^i dx^j \,,  
\label{PER_GEN2}
\ee
where the symbol $,_i$ denotes the derivative
with respect to the spatial coordinates.

If we work in the longitudinal gauge, the perturbed metric is:
\be
ds^2 = - (1 + 2 \Phi) dt^2 + a(t)^2 (1 - 2 \Phi) \delta_{ij}dx^i dx^j\, ,
\ee
where the two longitudinal perturbations are the same since neither 
the scalar field nor the perfect fluid have linear anisotropic terms 
in the pressure. The energy-momentum tensor for the fluid is:
\be
T_{\mu \nu} = (\rho + p) u_\mu u_\nu + p g_{\mu \nu} \,, \quad \quad 
u_\mu \, u^\mu = - 1 
\ee
and its perturbations are given by:
\be
\delta T^0_{{\rm F} \, 0} = - \delta \rho_F \,,  \quad \quad 
\delta T^0_{{\rm F} \, i} = \psi_{F \,, i} \quad \quad
\delta T^i_{{\rm F} \, j} = \delta p_F \, \delta^i_j 
\equiv w_F \delta \rho_F \, \delta^i_j \,,
\ee
where we have implicitly introduced the potential $\psi_F$ for the spatial 
velocity of the fluid $u_i$.

The perturbed total energy and pressure are, respectively:
\be
\delta\rho_{\rm tot} = \dot \phi \, \dot \delta\phi - \dot\phi^2 \Phi +
V_{\phi}\delta\phi + \delta \rho_F  \,, \label{deltarhotot} \ee
\be
\delta p_{\rm tot} = \dot \phi \, \dot \delta\phi - \dot\phi^2 \Phi -
V_ {\phi}\delta\phi + \omega_F \delta\rho_F \,. \label{deltaptot} \ee
The Einstein constraints are
\be
3 H \dot\Phi + 3 H^2 \Phi + \frac{k^2}{a^2} \Phi = - 4\pi G \delta
\rho_{\rm tot}
\ee
\be
 H \Phi + \dot\Phi = 4\pi G \left( \dot \phi \delta \phi - \psi_F \right) 
\label{g0i}
\ee
and the conservation of the energy momentum tensor for the fluid leads to:
\begin{eqnarray}
\delta\dot\rho_F + 3 H (1 + \omega_F)\, \delta \rho_F -
\frac{k^2}{a^2}\psi_F - 3 (1 + \omega_F) \rho_F \dot\Phi &=&
X_F \Phi + \delta X_F 
\nonumber \\
&=& 2 \Gamma \,\dot\phi\,\delta\dot\phi - \Gamma \,\dot\phi^2 \,\Phi
+ \delta\Gamma \,\dot\phi^2 \,,
\label{deltaF}
\end{eqnarray}
\begin{eqnarray}
\dot\psi_F + 3 H \psi_F + \Phi (1 + \omega_F) \rho_F +
\omega_F \delta \rho_F &=&
- \Gamma \dot\phi \delta\phi \,,
\label{deltapsi}
\end{eqnarray}
with
\begin{eqnarray}
\delta X_F &=& 2 \Gamma \dot \phi \dot{\delta \phi} - 2 \Gamma \dot \phi^2 \Phi
+ \dot \phi^2 \delta \Gamma = - \delta X_\phi \,.
\end{eqnarray}

The equation of motion for field fluctuations is:
\begin{eqnarray}
\ddot{\delta\phi} + (3H + \Gamma) \,\dot{\delta\phi} + \left[ \frac{k^2}{a^2} +
  V_{\phi\phi} \right] \delta\phi  &=&
4\, \dot\phi\, \dot \Phi - 2 V_{\phi} \Phi - \Gamma \dot \phi \Phi - 
\dot\phi\delta \Gamma
\label{deltaphi}
\end{eqnarray}

The curvature perturbation in the longitudinal gauge $\zeta$ is defined as:
\be \zeta = \Phi - \frac{H}{\dot H}\left(\dot\Phi + H \Phi\right) = \Phi +
\frac{2}{3} \frac{\dot\Phi +  H \Phi}{H (1 + \omega_{\rm tot})}. \label{zeta}
\ee
Following \cite{Bartolo-Corasaniti} we define the entropy perturbation between the scalar field and the fluid:
\be
 S_{\phi F} \equiv \frac{3H \gamma_F \gamma_{\phi} \rho_F}{\gamma_{\rm tot} \rho_{\rm tot}} \left(\frac{\delta\rho_{\phi}}{\dot\rho_{\phi} }
   -\frac{\delta\rho_F}{ \dot\rho_F} \right) \, =
 \,  \frac{\gamma_F \gamma_{\phi}
   \rho_F}{\gamma_{\rm tot}\rho_{\rm tot}} \left[\frac{\delta\rho_F}{
     \gamma_F \rho_F - r \gamma_{\phi }\rho_{\phi}} -
     \frac{\delta\rho_{\phi}}{ \gamma_{\phi}\rho_{\phi}(1 + r) } \right]
\label{entropy}
\ee
and the intrinsic entropy perturbation of the scalar field:
\be  S_{\phi} \equiv \frac{3H \gamma_{\phi} c^2_{\phi}} {1 - c^2_{\phi} }
\left(\frac{ \delta\rho_{\phi}}{ \dot\rho_{\phi}} -\frac{ \delta
    p_{\phi}}{\dot\rho_{\phi}}  \right) \, = \, \frac{\delta p_{\phi} -
  c^2_{\phi}  \delta\rho_{\phi} }{ \rho_{\phi} (1 - c^2_{\phi})(1 + r)} \, .
\label{entropyfield} \ee
We conclude this section with the equation of evolution for $\zeta$:
\be \dot\zeta = \frac{2}{3 H \gamma_{\rm tot}}\left[ - c_s^2
  \frac{k^2}{a^2}\Phi + 4\pi G \delta p_{\rm nad} \right] \, ,\label{dotzeta} \ee
where $\delta p_{\rm nad} \equiv \delta p_{\rm tot} - c_s^2 \delta \rho_{\rm
  tot}$ is the non-adiabatic pressure which depends on the intrinsic and
relative entropy perturbations, as we can see making use of
Eq. (\ref{soundtot}):
\be \frac{\delta p_{\rm nad}} {\rho_{\phi}(1 + r)} = (1 -  c^2_{\phi}) S_{\phi}
+ (\omega_F - c^2_{\phi})( 1 - \frac{\gamma_{\phi}\rho_{\phi}}{\gamma_F\rho_F}
r) S_{F \phi}.
 \label{pnad} \ee
It is simple to see from Eqs.(\ref{dotzeta}) and (\ref{pnad}) that 
in the case of pure radiation ($ \phi = \dot \phi = \delta \phi = 0$ 
and $ \omega_F = c_s^2 = 1 / 3$) curvature perturbations are conserved
on large scales.


\vspace{.5cm}

{\bf Large Scale Curvature Solution during Slow-Roll}

\vspace{.5cm}

It is useful to give the analytic approximation of the adiabatic solution 
for large-scale cosmological 
perturbation during slow-roll. By using 
Eqs. (\ref{slowphi},\ref{slowF},\ref{epsilon}), 
the approximate solutions at leading order 
in the slow-roll parameters are:
\ba
\Phi & \simeq &  \frac{4 \pi G}{H} \dot \phi (1 + r) \delta \phi   
\,\,\,\,\,\,\,\,\,
\nonumber \\
\psi_F & \simeq & - r \dot\phi \delta \phi \,,
\nonumber \\
\delta \rho_F & \simeq & - 
\frac{H\dot\phi}{(1+r)(1 + \omega_F)} \left(2\, 
\frac{V_{\phi \, \phi}}{3 H^2} \,r - \beta\,(1-r) -  
      (3+r) \,r \, \epsilon \right)\delta\phi\,\,\, \nonumber \\
\dot\delta\phi &\simeq& - \frac{H}{1+r}\left(\frac{V_{\phi \, \phi}}{3 H^2}
+ \beta - (2 + r) \epsilon \right)\delta\phi \,, \nonumber \\
\delta \phi &\propto& \frac{\dot \phi}{H} \,,
\label{approx}
\ea
with $\beta = \dot \Gamma / (3 H^2)$. 
These solutions have been obtained neglecting the highest time 
derivatives and the terms $\propto k^2$ 
in Eqs. (\ref{g0i},\ref{deltaF},\ref{deltapsi},\ref{deltaphi}) and are valid if 
$\Gamma = \Gamma ( \phi )$. 
It is interesting to use the above relations 
and write the leading term (in slow-roll parameter) of curvature perturbations:
\be
\zeta \simeq 
\frac{2}{3} \frac{\Phi}{1 + \omega_{\rm tot}} = 
\frac{\Phi}{\epsilon} \simeq \frac{H}{\dot \phi} \delta \phi \,,
\label{zeta_sr}
\ee 
which is formally just the single field expression; 
$\Gamma$ is however present in the dynamics for the 
inflaton, which differs from single field case with the same $V(\phi)$ and 
$\Gamma = 0$. The expression differ from the multi-field inflationary 
case in which all the scalar fields in slow-roll contribute (with the same 
formal weight) to curvature perturbations. We end on noting that $\zeta$ 
is constant in time for the last expression of Eq. (\ref{approx})
at leading order in slow-roll parameters 
(i.e. $\dot \zeta \sim H \zeta {\cal O}(\epsilon)$).


\vspace{0.2cm}

\section{Numerical Evolution of Cosmological Perturbations}

We will consider the numerical evolution of linear 
cosmological perturbations from inflation through reheating, following 
Ref. \cite{FB_PRD} which evolved the Mukhanov variables related to two 
scalar fields in a fully constrained evolution. 

In this case we need to obtain the evolution of the Mukhanov variable 
associated to the scalar field coupled with Lukash variable associated to the 
gradient velocity of the perfect fluid. Given a fully general metric 
perturbation as in Eq. (\ref{PER_GEN2}), gauge invariant variables for the scalar 
field and the fluid are defined as:
\be
\delta \phi^{\rm g.i.} = 
\delta \phi + \frac{\dot \phi}{H} \psi \quad \quad 
\psi^{\rm g.i.}_F = \psi_F - (\rho + p) \psi,_i \,.
\label{gi}
\ee
In the UCG, where the spatial metric is unperturbed,
\be
ds^2=-(1+2 \alpha)dt^2 - a \beta_{,i} dt dx^i+a^2 \delta_{ij} dx^i dx^j
\,,
\label{PER_UCG2}
\ee
the gauge invariant variables coincide with the scalar field and 
fluid velocity potential fluctations.
It is therefore easier to obtain the evolution for these gauge invariant 
variables in such a gauge. From here on we restrict ourselves to this gauge and we 
shall write fluctuations of the scalar field and of the fluid, although 
different, in the same way as the previous section.


The equations corresponding to (\ref{deltaF},\ref{deltapsi},\ref{deltaphi}) 
are:
\begin{eqnarray}
\delta\dot\rho_F + 3 H (1 + \omega_F)\, \delta \rho_F -
\frac{k^2}{a^2}\psi_F - (1 + \omega_F) \rho_F \frac{k^2}{2 a} \beta
&=& X_F \alpha + \delta X_F
\nonumber \\
&=& 2 \Gamma \,\dot\phi\,
\delta\dot\phi -  \Gamma \,\dot\phi^2 \,\alpha + \delta\Gamma \,\dot\phi^2
\label{deltaF_UCG}
\end{eqnarray}
\be
\dot\psi_F + 3 H \psi_F + \alpha (1 + \omega_F) \rho_F
+ \omega_F \delta \rho_F =
- \Gamma \dot\phi \delta\phi.
\label{deltapsi_UCG}
\ee
\begin{eqnarray}
\ddot{\delta\phi} + (3H + \Gamma)
\,\dot{\delta\phi} + \left[ \frac{k^2}{a^2} +
  V_{\phi\phi} \right] \delta\phi  &=& 
\dot \phi \left( \dot \alpha + \frac{k^2}{2 a} \beta \right)
- 2 V_\phi \alpha - \dot\phi\delta\Gamma - \alpha \Gamma \dot \phi\,.
\label{deltaphi_UCG}
\end{eqnarray}

If $\delta \phi$ is used to quantize the linear scalar field
fluctuation in presence of gravity \cite{mukhanov}, the
variable to quantize a fluid \cite{lukash}, 
as also explained in \cite{MFB}, is:
\be
Q_F \equiv \frac{\psi_F}{\sqrt{\rho_F (1 + w_F)}} \,.
\ee
On restricting to the case in which $\Gamma$ is constant in time, 
the coupled second order differential equation for
$(Q_\phi \,, Q_F) = (\delta \phi \,, Q_F)$ are:
\be
{\ddot Q_i} + \left( 3 H \, \delta_{i j} + G_{i j} \right) \, {\dot Q_j} +
{\Omega_{i j}} \, Q_j =0
\label{system}
\ee
where $G$ has the following components
\begin{eqnarray}
G_{\phi\, \phi} &=& \Gamma \\
G_{\phi \,F} &=& 4 \pi G (w_F^2 -1)\rho_F \dot \phi \over \sqrt{\rho_F (1 + w_F)} w_F  H \\
G_{F \,\phi} &=& {\left[ -4 \pi G (w_F^2 -1)\rho_F + (2 w_F +1) H \Gamma \right] \dot \phi \over \sqrt{\rho_F (1 + w_F)} H} \\
G_{F \, F} &=& \Gamma {{\dot\phi}^2 \over \rho_F}
\, ,
\end{eqnarray}
and where $\Omega$ is given by
\begin{eqnarray}
\Omega_{\phi \, \phi} &=&
{k^2 \over a^2} + V_{\phi \phi} 
+ 16 \pi G {\dot\phi \, V_{\phi}\over H}
+ 24 \pi G {\dot\phi}^2 
+ 4 \pi G \left( {2 \,w_F -1 \over w_F}\right){{\dot\phi}^2 \, \Gamma \over H}
\nonumber \\
& & - 16 \pi^2 G^2 {{\dot\phi}^2\over H^2} \left(\rho_F  
{(w_F +1)^2 \over w_F} + 2 {\dot \phi}^2  \right) \\
\Omega_{\phi \, F} &=&
\sqrt{\rho_F (1 + w_F)}
\left( -8 \pi G {V_{\phi} \over H}
- 6 \pi G {(1 + w_F)^2\over w_F} {\dot \phi}
+ 16 \pi^2 G^2 {(1 + w_F)^2\over w_F} \rho_F {{\dot \phi} \over H^2}
\right. \nonumber \\
& & \left. 
+ 32 \pi^2 G^2 {{\dot \phi}^3 \over H^2}
+ 2 \pi G {(w_F-1)\over w_F} {\Gamma \over H} {{\dot \phi}^3 \over \rho_F}
- 4 \pi G \Gamma \frac{\dot \phi}{H}\right)\\
\Omega_{F \, \phi} &=&
\sqrt{\rho_F (1 + w_F)}
\left( -4 (1 + w_F) \pi G {V_{\phi} \over H}
- 12 \pi G (1+w_F) {\dot \phi}
- {\Gamma V_{\phi} \over (1+w_F) \rho_F}
+ {\left( 3 w_F H -\Gamma \right) \Gamma {\dot \phi} \over (1+w_F) \rho_F}
\right. \nonumber \\
& & \left. + 32 \pi^2 G^2 (1+w_F) {\rho_F {\dot \phi} \over H^2}  
+ 16 \pi^2 G^2 (1+w_F) {{\dot \phi}^3 \over H^2}
+ 4 \pi G {1 \over (w_F+1)}{\Gamma {\dot \phi}^3 \over H \rho_F}
\right)\\
\Omega_{F \, F} &=&
w_F{k^2 \over a^2}
+ {9 \over 4}H^2(1-w_F^2)
+ 6 \pi G \rho_F (1+w_F)(1+3w_F)
- {32 \pi^2 G^2 \rho_F^2 \over H^2}(1+w_F)^2
\nonumber \\
& &
- \Gamma {V_{\phi}{\dot\phi}\over \rho_F}
+ 6 \pi G {\dot \phi}^2 (w_F-1)
- {4 \pi G \Gamma {\dot \phi}^2\over H}
+ {\Gamma {\dot \phi}^2 \over 2 \rho_F}(- 2 \Gamma + 3 w_F H )
\nonumber \\
& &
- {16 \pi^2 G^2 \over H^2}\rho_F {\dot \phi}^2 (1+w_F)^2
-{{\Gamma}^2{\dot \phi}^4 \over {4 \rho_F^2}}
\, .
\end{eqnarray}

Several cross-checks on the novel set of equations can be made:

1) For $\rho_F = \Gamma = 0$ the equation for $Q_\phi$ agrees with
the Mukhanov equation for single field inflation \cite{mukhanov}.

2) When the two components are decoupled ($\Gamma=0$) and the fluid is stiff
matter ($w_F=1$) the above equations agree with those for two scalar fields
\cite{FB_PRD} when the second one $\chi$ is decoupled and massless
(i.e. equivalent
to stiff matter) reminding that $Q_\chi = - Q_F$.

We now evolve the system in Eq. (\ref{system}) starting from wavelength well inside the Hubble radius 
$k >> a H$ with initial conditions:
\be
Q_\phi \simeq \frac{e^{-i k \eta}}{a (2 k)^{1/2}} \,, 
\quad \quad Q_F \simeq \frac{e^{-i w_F^{1/2} k \eta}}{a (4 k^2 w_F)^{1/4}} \,.
\label{ic}
\ee

The gauge-invariant curvature perturbation $\zeta$ 
in UCG gauge is written as:
\be
\zeta = H 
\frac{\dot \phi \, Q_\phi - \sqrt{\rho_F (1 + w_F)} \, Q_F}{{\dot \phi}^2 
+ \rho_F ( 1 + w_F) } \equiv \zeta_\phi + \zeta_F \,.
\ee

\begin{figure}
\begin{tabular}{cc}
\includegraphics[scale=0.71]{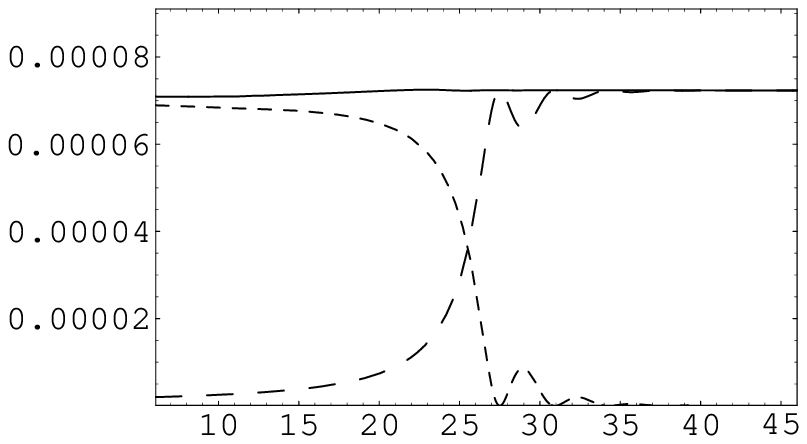}
\includegraphics[scale=0.71]{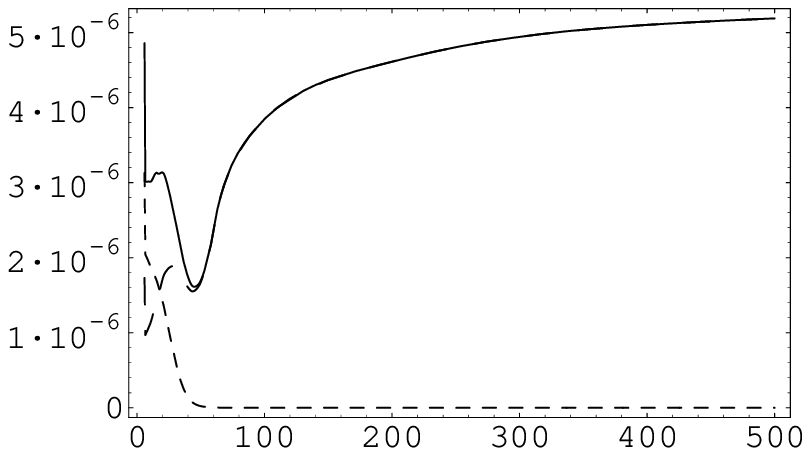}
\includegraphics[scale=0.71]{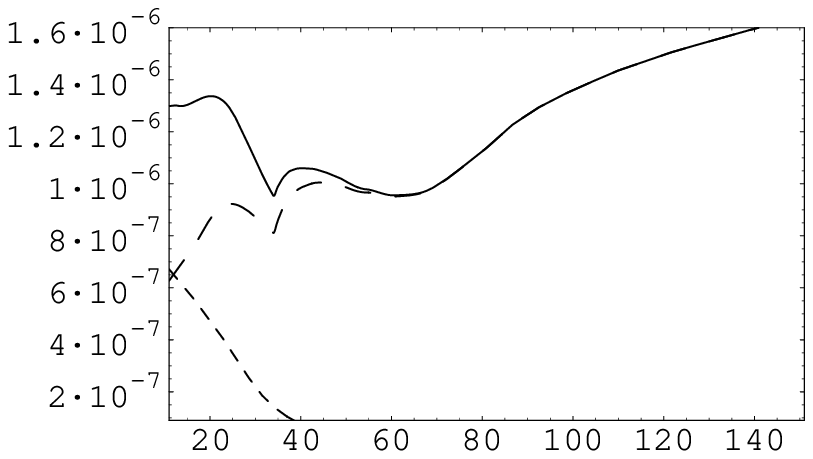}
\end{tabular}
\caption{Typical evolution of curvature perturbations 
(total as solid line, short-dashed and long-dashed for the field 
$\zeta_\phi$ and fluid $\zeta_F$ contribution, respectively)  
from inflation to radiation as function of an adimensional
cosmic time $m \, t$.
On the left panel, the evolution for $\Gamma /m = 0.5$ for which
the curvature perturbations are given by the inflaton
(solid and short-dashed lines are super-imposed) and there is negligible
post-inflationary change.
On the middle and right panel, the evolution for
$\Gamma/m = 6$ and $\Gamma/m = 8$.
}
\label{fig2}
\end{figure}

Evolution of curvature perturbations are displayed in Fig.
\ref{fig2}. Perfect fluid
fluctuations are amplified similarly to inflaton fluctuations when
$\Gamma \ne 0$, leading to a mixture of curvature and isocurvature
fluctuations during slow-roll. Isocurvature
perturbations can then change the amplitude curvature perturbations during
and after the inflaton decay.


\vspace{.5cm}


{\bf The uncoupled case}

\vspace{.5cm}

As is clear from the numerics, the perfect fluid does not contribute to 
curvature perturbations in the uncoupled case $\Gamma=0$. We have already 
observed that for $w_F > 0$, the homogeneous perfect fluid energy density 
is rapidly washed out for $\Gamma = 0$. If we neglect the terms which 
decay quasi-exponentially on time, the equation for fluid perturbations 
becomes: 
\be
{\ddot Q_F} + 3 H {\dot Q_F} + \left[ w_F{k^2 \over a^2}
+ {9 \over 4}H^2(1-w_F^2)
+ 6 \pi G {\dot \phi}^2 (w_F-1)
\right] Q_F \simeq 0 \,.
\label{qf}
\ee
For radiation, at leading order in a nearly de Sitter exponent, 
we have an effective mass $\simeq 2 H^2$ which makes the fluid fluctuations 
similar to a massless conformally coupled fluid. The initial vacuum fluid 
fluctuations in Eq. (\ref{ic}) are almost left unchanged during the 
streching to large scale, leading to a spectrum 
$P_F (k) = k^3 |Q_F|^2/(2 \pi^2) \sim k^2$, which is too blue to affect 
the nearly scale-invariant spectrum of inflaton fluctuations in the 
observable range.


\section{Discussions and Conclusions}

We have studied the coupling of the inflaton to a perfect fluid through 
a friction term $\Gamma$ during and after inflation. 
Although this type of coupling 
has been used in the early days to describe the decay of the 
inflaton after the accelerated era, 
it has also been used in the regime $\Gamma \gg H$ for warm 
inflation.

By considering radiation as the perfect fluid and $\Gamma$ 
constant in time for simplicity, we have shown how 
this coupling freezes the perfect fluid during 
the slow-roll evolution and large-scale fluid  
fluctuations are amplified together with the scalar field ones, leading to 
a mixture of curvature and isocurvature 
fluctuations during the slow-roll regime.
In the stage of the inflaton decay, 
large scale curvature and isocurvature perturbations are not weakly coupled 
anymore and there is a transfer of the latter leading to an amplitude 
for the former which is different from the one computed during 
slow-roll. After the decay is completed, a purely adiabatic curvature 
perturbation is left.
We believe that the qualitative behaviour found here is generic and 
does not occur only for the simplest parameters studied in this paper - 
radiation as the perfect fluid and $\Gamma$ constant in time. 
Note that a variation of large scale curvature 
perturbations was also found in the late decay of a massive curvaton \cite{MUW}.

Our results for a coupling to a perfect fluid add to the changes of 
curvature perturbations during preheating 
found for other couplings to scalar fields, as studied in \cite{FB_PRD}.   
As already said, this coupling of the inflaton easily generates  
non-negligible isocurvature fluctuations during inflation, which is one of 
the key requirements in order to have a change of curvature perturbations in 
the post-inflationary stage \cite{FB_PRD}. Since we see that the amount of 
isocurvature perturbations increases with $\Gamma$ in the cases studied 
here, inflationary models with sizeable $r$ have a larger variation of 
curvature perturbations in the post-inflationary era. As shown explicitly 
here, a successfull conversion of isocurvature into curvature perturbations 
at reheating may not need a varying decay rate.

\vspace{1cm}

{\bf Acknowledgements}

We wish to thank Robert Brandenberger and 
Alessandro Cerioni for comments on the manuscript.






\end{document}